%% file: Example.tex
\documentclass[a4paper,twoside]{article}
\usepackage{epsfig}
\usepackage{subcaption}
\usepackage{calc}
\usepackage{amssymb}
\usepackage{amstext}
\usepackage{amsmath}
\usepackage{amsthm}
\usepackage{multicol}
\usepackage{pslatex}
\usepackage{apalike}
\usepackage{wrapfig}
\usepackage{flushend}
\usepackage{lipsum}
\usepackage{soul,color}
\usepackage{empheq}
\usepackage{tikz}
\usepackage{pgf}
\usepackage{pgfplots}
\usepackage{microtype}
\usepackage{SCITEPRESS}     % Please add other packages that you may need BEFORE the SCITEPRESS.sty package.

\input{mathCommands}
\newtheorem{remark}{Remark}
\newtheorem{ass}{Assumption}
\newtheorem{definition}{Definition}

\newtheorem{lemma}{Lemma}
\newtheorem{thm}{Theorem}

\def\-{\raisebox{.75pt}{-}}
\def\+{\raisebox{.75pt}{+}}
\begin{document}

\title{Lyapunov Stability of a Nonlinear Bio-inspired System for the Control of Humanoid Balance.}

\author{\authorname{Vittorio Lippi\sup{1}, Fabio Molinari\sup{1}}
\affiliation{\sup{1}Technische Universit\"{a}t Berlin, Fachgebiet Regelungssysteme, Einsteinufer 17 D-10587, Berlin, Germany}
\email{  \{vittorio.lippi$\vert$molinari\}  @tu-berlin.de}}

 %more to come

\abstract{Human posture control models are used to analyse 
	neurological experiments and 
	control of humanoid robots. This work focuses on a well-known nonlinear posture control model, the DEC (Disturbance estimate and Compensation). 
	In order to compensate disturbances,
unlike other models, %for compensation of disturbance,
DEC feedbacks signals coming from sensor fusion rather than \textit{raw} sensory signals. 
In previous works, the DEC model is shown to predict human behavior
and to provide a control system for humanoids. 
In this work, the stability of the system in the sense of Lyapunov is formally analysed. 
The theoretical findings are
combined with simulation results, in which an external perturbation 
of the support surface reproduces a typical 
scenario in posture control experiments.}

\onecolumn \maketitle \normalsize \setcounter{footnote}{0} \vfill

\section{\uppercase{Introduction}}
Mathematical models of human balance are used for the analysis of neurological experiments \cite{vanderKooij2007,van2005comparison,van2006disentangling,goodworth2018identifying,mergner2010,engelhart2014impaired,pasma2014impaired,jeka2010dynamics,boonstra2014balance}, and for the control of humanoid robots.  Most of human posture control studies exploit linear models such as the \textit{independent channel} model \cite{peterka2002sensorimotor}, 
that assumes a linear and time invariant behaviour \cite{engelhart2016comparison}. Linear models have the advantage of being simple to analyse and relatively easy to be fit on data. 
However, 
experiments reveal that human posture control exhibits important non-linearities. 
\newline
In this work, we study the stability of a non-linear bio-inspired posture control system, the DEC, \textit{Disturbance estimate and Compensation} \cite{mergner2010}. 
\begin{figure}[t!]
	\centering
		\includegraphics[width=1.00\columnwidth]{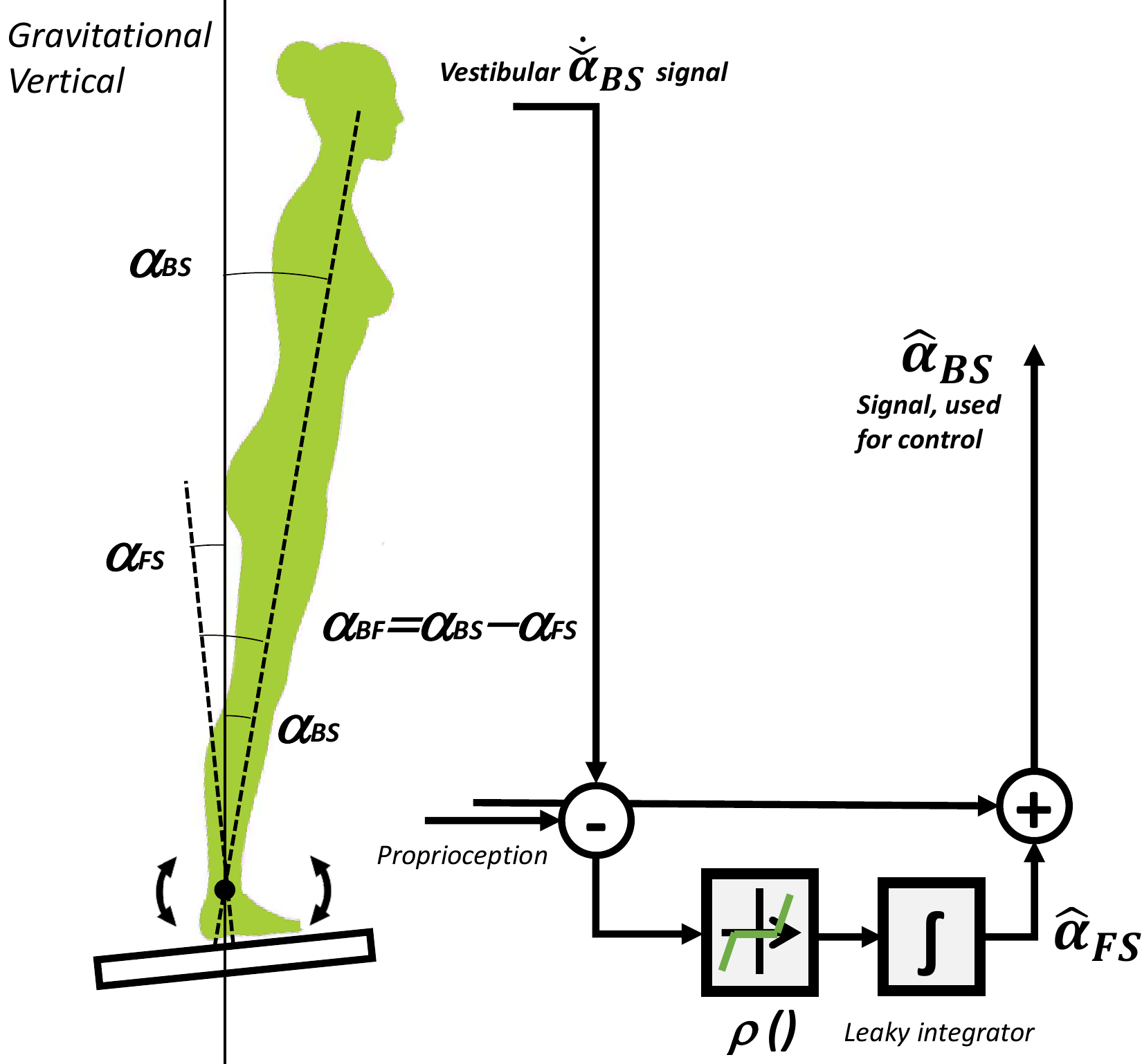}
	\caption{Posture control model. On the left: illustration of the scenario and definition of angles used in text. On the right: schema of the bio-inspired sensor fusion.}
	\label{fig:Fig1}
\end{figure}
The DEC model consists of a servo control loop and a compensation of external disturbances estimated on the basis of sensory inputs. The control principle can be addressed as ``feed forward disturbance correction'' \cite{luecke1968analysis,roffel2007process,zhong2012feedforward} or, in German, ``St{\"o}rgr{\"o}{\ss}enaufschaltung'' \cite{bleisteiner2013handbuch}.  Throughout this paper, the DEC is used to model a scenario 
where the subject stands on a tilting surface. 
We analyse the effect of a dead-band nonlinearity that affects the 
sensory-based estimate of the support surface tilt. 
Such nonlinearity, common in literature,
is assumed on the basis of the behaviour observed in humans. 
The formal conditions for Lyapunov stability are investigated.
\newline
The paper is organized as follows: in Section~\ref{problemDescription},
the control problem is introduced and the body mechanics is described; 
Section~\ref{humansense} provides details about human-inspired sensor fusion and actuation; 
the conditions for stability are obtained in Section~\ref{stability},
where evidence is also provided; 
Section~\ref{sim} presents simulations
and a qualitative discussion of the system behaviour; 
conclusions and future work are presented in Section~\ref{conclusions}.

\label{dynamicSys}
\section{Problem Description}
\label{problemDescription}
%\subsection{Body mechanics}
\label{bodymechanics}
\begin{figure}
	\centering
\includegraphics[width=1.00\columnwidth]{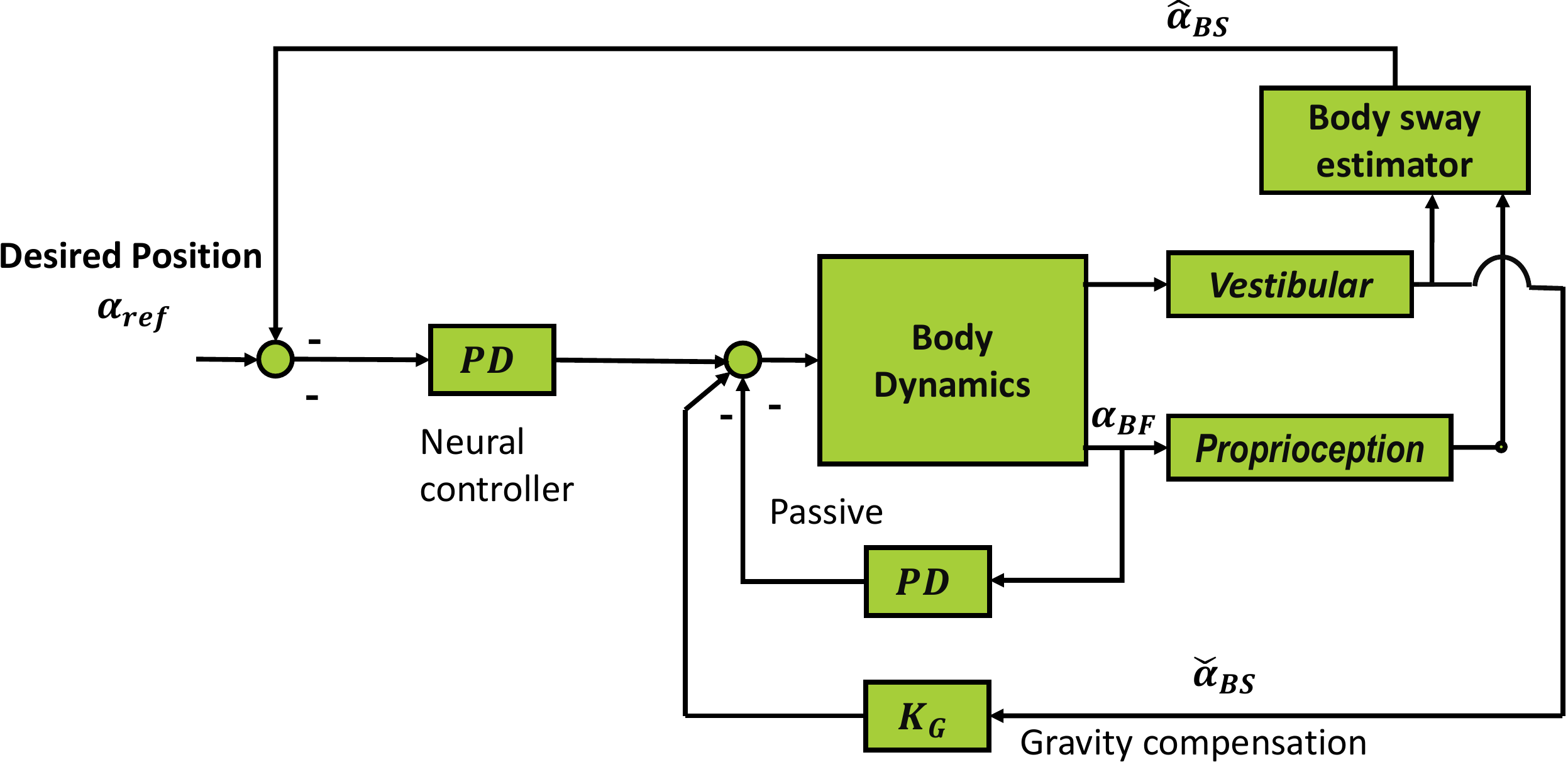}

	\caption{A scheme of a controller based on the DEC concept. The sensory inputs are used to reconstruct the physical disturbances acting on the body (in this scenario gravity and support surface tilt). The system is controlled by a servo controller, consisting of a PD regulator setting the desired body position, and a direct compensation of the estimated gravity disturbance.}
	\label{fig:DECscheme}
\end{figure}
\input{bodyMechanics}

\begin{table*}
	\begin{tabular*}{\textwidth}{|c|c| @{\extracolsep{\fill}}c |}
		\hline
		\textbf{Variable}&\textbf{Defined in}&\textbf{Definition}
		\\
		\hline
		$\alpha_{BF}$ & (\ref{eq:angles_eq}) & Ankle joint angle
		\\
		\hline
		$\alpha_{BS}$ & (\ref{eqd1}) & Body sway respect to the vertical
		\\
		\hline
		$\alpha_{FS}$ & \S \ref{bodymechanics} & Support surface rotation
		\\
		\hline
		$\hat{\alpha}_{FS}$ & (\ref{eq:leakyInt})  & 
		\mbox{~Support surface rotation estimate based on sensor fusion (vestibular+proprioceptive)}
		\\
		%\hline
		%$\rho(\alpha)$ & eq. \eqref{deadband}  & deadband nonlinearity
		%\\
		%\hline
		%$\theta$ & eq. \eqref{deadband}  & threshold for the deadband nonlinearity
		%\\
		\hline
		$\hat{\alpha}_{BS}$ &  \eqref{eq:gettingalfahat} & Body sway estimate based on sensory input (vestibular+proprioceptive)
		\\
		\hline
		$\check{\alpha}_{BS}$ & \S \ref{sec:sensors} & Body sway estimate based on vestibular input
		\\
		%\hline
		%$K_p^p$ & \S \ref{bodymechanics} & ankle joint passive stiffness 
		%\\
		%\hline
		%$K_d^p$ & \S \ref{bodymechanics} & ankle joint passive damping 
		%\\
		\hline
	\end{tabular*}
	\caption{List of variables and their definition. The second column contains equation numbers, in parentheses, 
		or section number depending on where the variable is defined.}
\end{table*}
%%% 
\vspace{-2mm}
\section{Human-inspired Sensors and Actuation}
\label{humansense}
\input{measuredSignals}
\input{humanInspiredSFC}

\section{Stability Analysis}
\label{stability}
\input{system}

\section{Simulation}
\label{sim}
\begin{figure}[t]
	\includegraphics[width=\columnwidth]{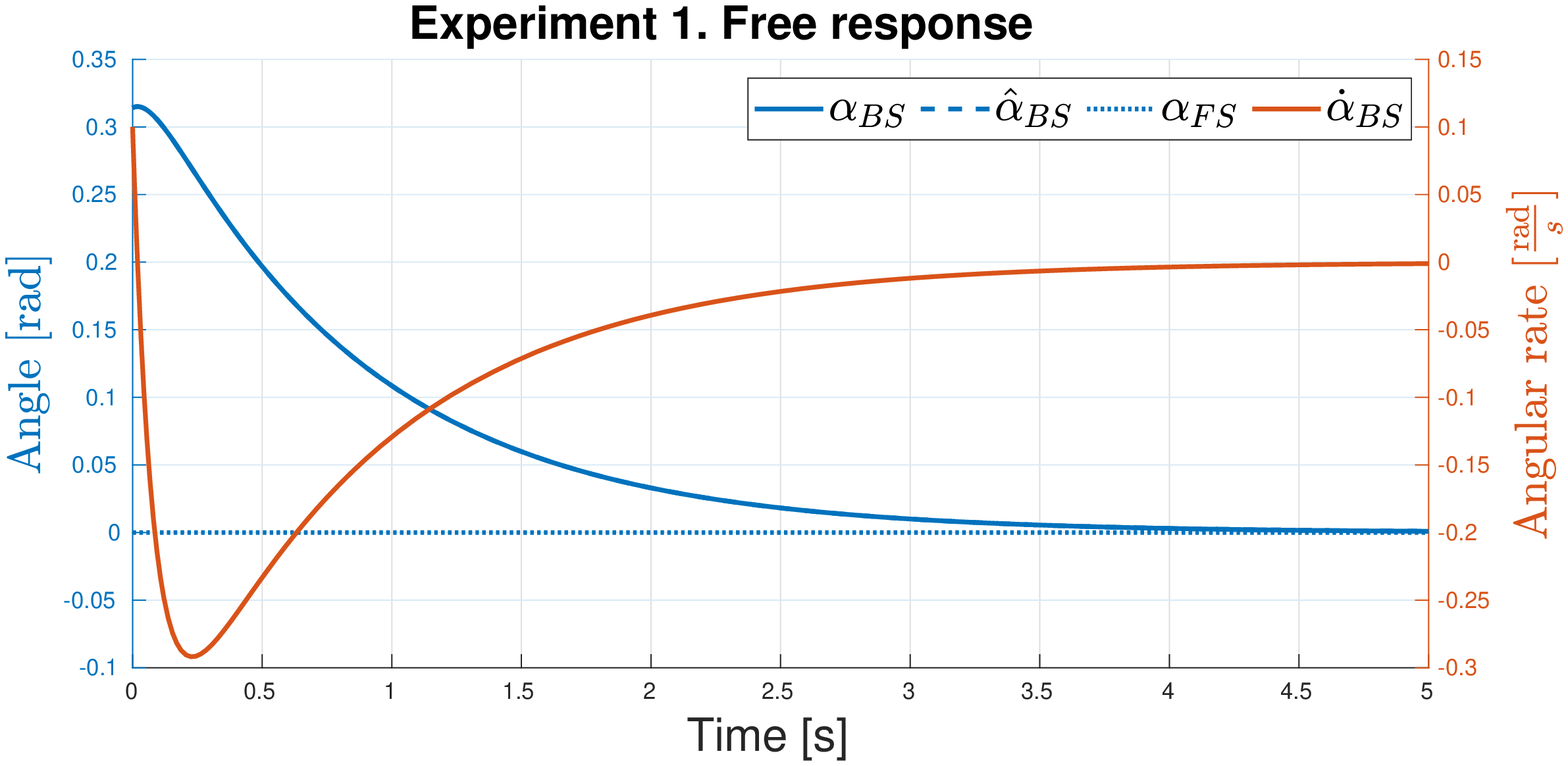}
	\caption{Free response with horizontal support surface and initial conditions $\mathbf{x}(0)=[\pi/10, 0.1, \pi/10, 0]^T.$ The estimate $\hat{\alpha}_{BS}$ is equivalent to $\alpha_{BS}$ when $\alpha_{FS}=0$.}
	\label{fig:Exp1}
\end{figure}

\begin{figure}[t]
	\includegraphics[width=\columnwidth]{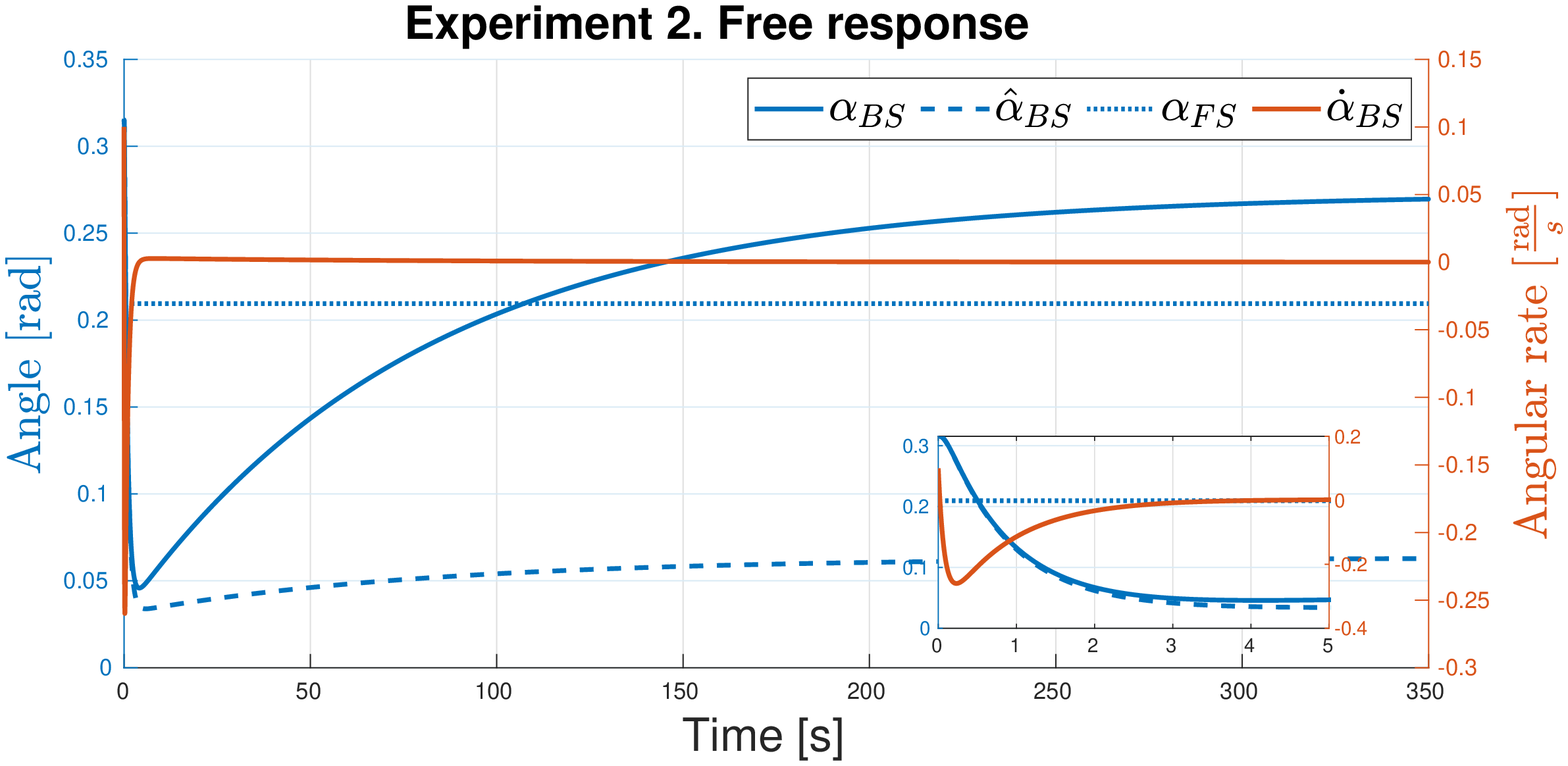}
	\caption{Free response with tilted, but not moving, support surface and non-zero initial body sway velocity and body lean, i.e. $\mathbf{x}(0)=[\pi/10, 0.1, \pi/10, \pi/15]^T.$ The smaller plot shows the transient during the first 5 seconds.}
	\label{fig:Exp2}
\end{figure}

\begin{figure}[t]
	\includegraphics[width=\columnwidth]{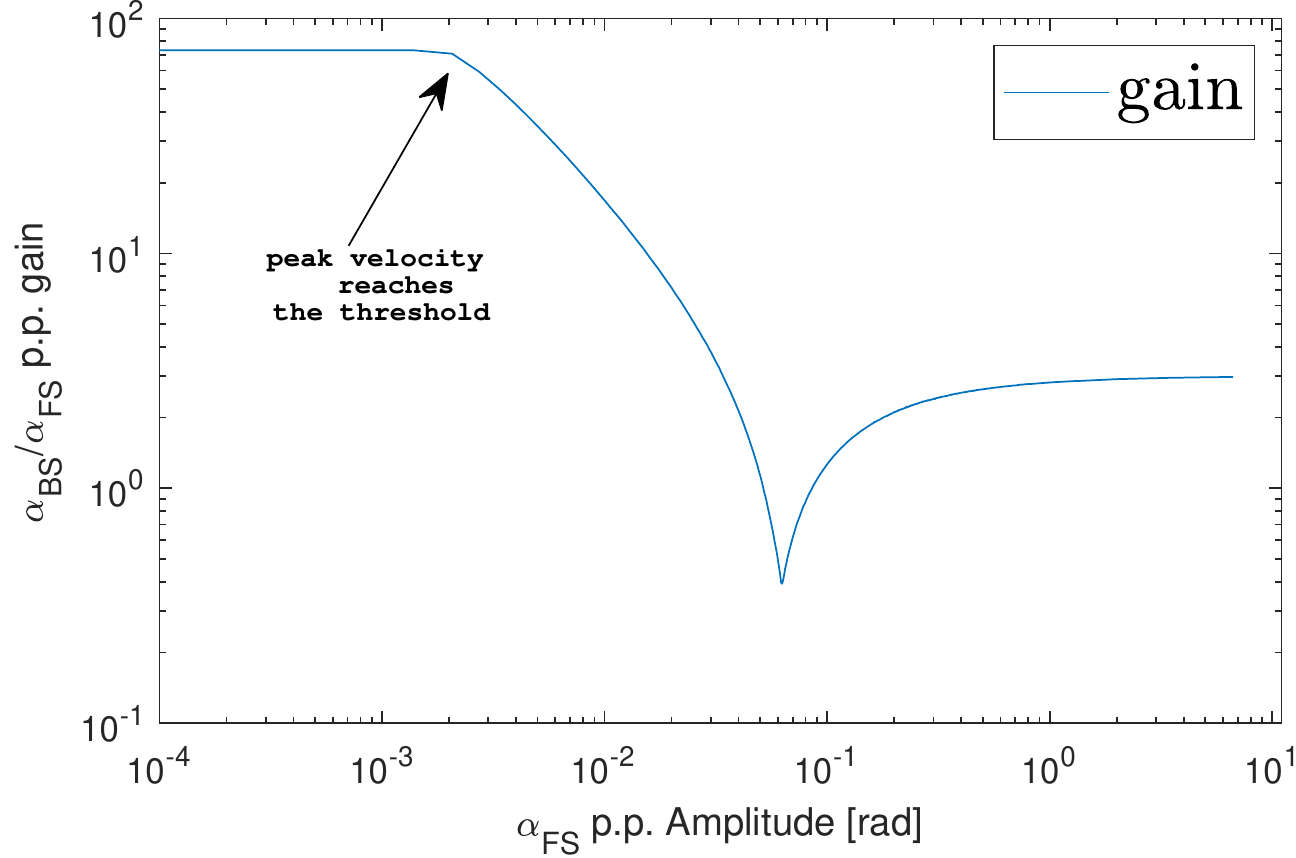}
\caption{Support surface tilt to body sway gain with sinusoidal support tilt at different amplitude. The gain is computed as the ratio between peak to peak amplitude for of the input and the one of the output. Smaller support surface tilt are associated with larger gains because they are under-compensated due to the nonlinearity. The plateau on the left is the zone of linear behavior that happens when the support surface rotation speed is always under threshold $\theta$. For larger amplitudes the gain tends asymptotically to a constant gain, i.e. linear behavior}
\label{fig:nonlin}
\end{figure}

The parameters,
defined on the basis of human anthropometrics 
(see \cite{winter2009biomechanics}) 
and previous posture control analysis (see \cite{T.Mergner2009,Mergner2003,G.Hettich2014}), 
are shown in Table \ref{tab:SystemParameters}. 
\begin{table}[htb]
	\centering
		\begin{tabular}{|l|l|}
			\hline
			Parameter & Value \\
			\hline
			\hline
			$J_B$ & $71.55 \ \mathrm{Kg} \cdot \mathrm{m}^2$\\
			\hline
			$K_g$ & $0.8 $ \\
			\hline
			$K_p^p$ & $157.31 \ N \cdot \mathrm{m}$\\
			\hline
			$K_d^p$ & $ 39.32 \ N \cdot \mathrm{m} \cdot s$\\
%			\hline
%			$K_p^a$ & $629.24.77 \ N \cdot m$\\
%			\hline
%			$K_d^a$ & $157.31 \ N \cdot m \cdot s$\\
			\hline
      $c_L$ & $0.0125 \ s^{-1}$ \\
			\hline
			$\theta$ & $0.0028 \ rad$ \\
			\hline
			$m$ & $80~\mathrm{Kg}$ \\
			\hline
			$h$ & $1.80~\mathrm{m}$ \\
      \hline
		\end{tabular}
	\caption{System parameters}
	\label{tab:SystemParameters}
\end{table}
With the specific set of parameters, 
and by (\ref{ineq1})-(\ref{ineq3}),
we design
$$
	K_p^a = -1200~N \cdot \mathrm{m}
$$
and
$$
	K_d^a = -1000~N \cdot \mathrm{m\cdot s}
.
$$
The behavior of the system is shown in regime of free response with no support surface tilt velocity, and forced response with a periodic input. Specifically the following conditions are simulated:\\
%\begin{itemize}
	\textbf{Condition 1}: free response with $$\mathbf{x}(0)=[\pi/10, 0.1, \pi/10, 0]^T.$$
	The free response with no support surface tilt is the characteristic one of a linear second-order system (see Fig. \ref{fig:Exp1}). This happens because the nonlinearity affects 
	only the input $u(t)$. The leaky integrator used in the estimate of $\alpha_{FS}$ is constantly at zero. \\ 
	\textbf{Condition 2}: free response with $$\mathbf{x}(0)=[\pi/10, 0.1, \pi/10, \pi/15]^T.$$
	The free response with a constant support surface tilt is shown in Fig. \ref{fig:Exp2}.
	The response is again the characteristic of a linear system, 
	in which $x_4$ behaves as a constant signal
	affecting the dynamics of $x_2$ and $x_3$.
%	in this case including a third pole 
%	(integrator).
%	due to the dynamics of the leaky integrator. 
	There is a residual lean $\alpha_{BS}$ due to the error in body sway estimate $\hat{\alpha}_{BS}$. \\
	\textbf{Condition 3}: forced response with $$\mathbf{x}(0)=[0, 0, 0, 0]^T$$ and $$u(t)=0.1\cos(10t) .$$ The forced response shows a partial rejection of the external disturbance. The effect of the nonlinearity is reflected in the difference between $\alpha_{BS}$ and its estimated value $\hat{\alpha}_{BS}$. The simulations is repeated with different amplitudes for the support surface tilt profile, producing the results in Fig. \ref{fig:nonlin}. The gain, in this context defined as the ratio between peak to peak amplitude for of the input and the output is plotted for different amplitudes. Smaller support surface tilt are associated with larger gains because they are under-compensated due to the nonlinearity. Specifically the plateau on the left is the zone of linear behavior that happens when the support surface rotation speed is always under threshold $\theta$ and the disturbance is not compensated. For larger amplitudes the gain tends asymptotically to a constant gain because the signal is almost always above the threshold. 
%\end{itemize}

%\textbf{Condition 1:} The free response with no support surface tilt is the characteristic one of a linear second-order system (see Fig. \ref{fig:Exp1}). This happens because the nonlinearity affects the input $u(t)$. The leaky integrator used in the estimate of $\alpha_{FS}$ is constantly at zero. 
%\textbf{Condition 2:}The Free response with a constant support surface tilt is shown in Fig. \ref{fig:Exp2}).The response is again the characteristic of a linear system, in this case including a third pole due to the dynamics of the leaky integrator. There is a residual lean $\alpha_{BS}$ due to the error in body sway estimate $\hat{\alpha}_{BS}$.
\begin{figure}[t]
	\includegraphics[width=\columnwidth]{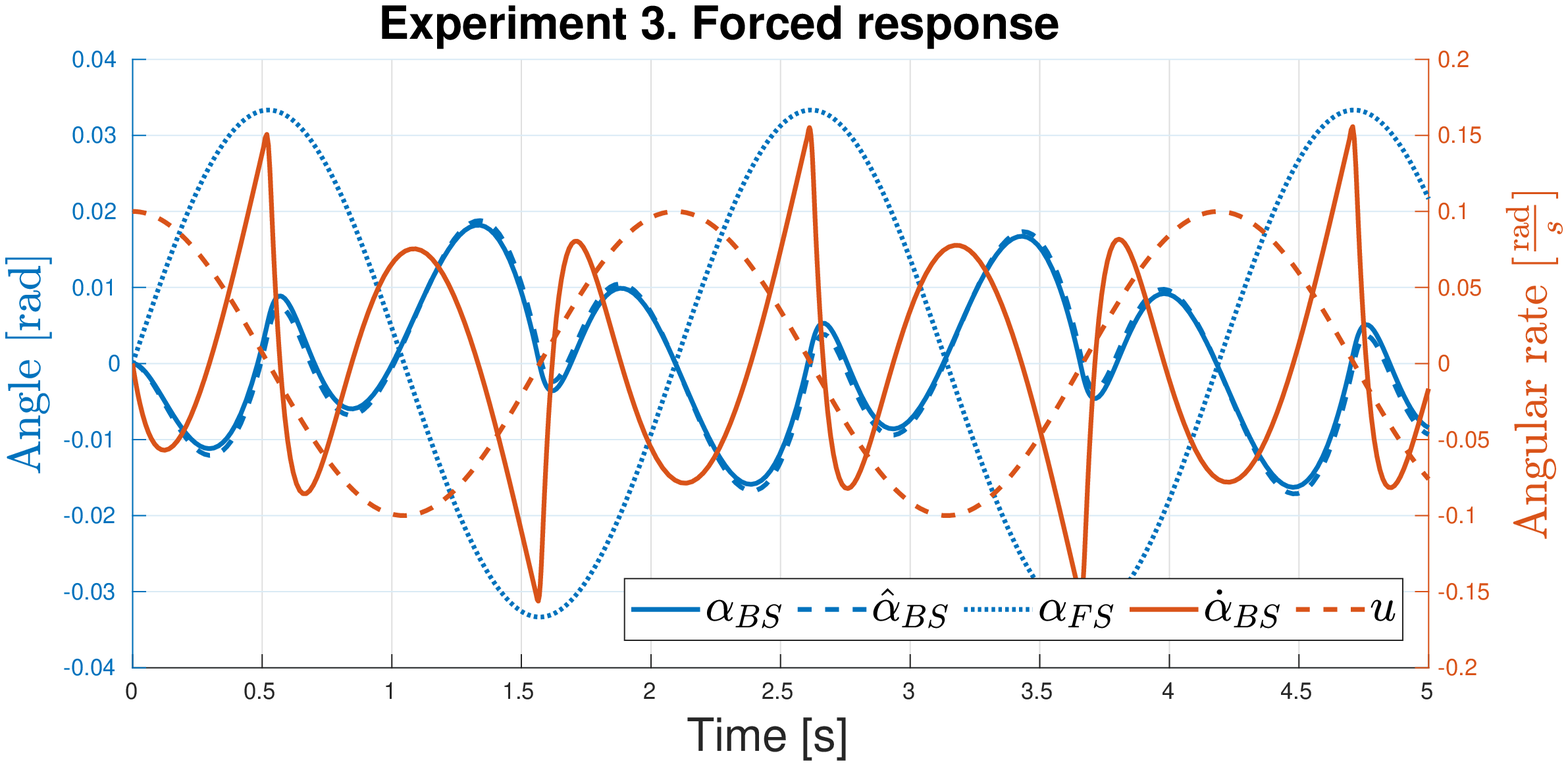}
	\caption{Forced response to sinusoidal support tilt with initial conditions $\mathbf{x}(0)=[0, 0, 0, 0]^T$, and input $u(t)=0.1\cos(10t).$}
	\label{fig:Exp3}
\end{figure}
%
%\textbf{Condition 3:} The forced response shows a partial rejection of the external disturbance. The effect of the nonlinearity is reflected in the difference between $\alpha_{BS}$ and its estimated value $\hat{\alpha}_{BS}$. The simulations is repeated with different amplitudes for the support surface tilt profile, producing the results in Fig. \ref{fig:nonlin}. The gain, in this context defined as the ratio between peak to peak amplitude for of the input and the output is plotted for different amplitudes. Smaller support surface tilt are associated with larger gains because they are under-compensated due to the nonlinearity. Specifically the plateau on the left is the zone of linear behavior that happens when the support surface rotation speed is always under threshold $\theta$ and the disturbance is not compensated. For larger amplitudes the gain tends asymptotically to a constant gain because the signal is almost always above the threshold. 
	
%
%the system from Eq. \ref{system} becomes
%\footnotesize
%\begin{equation}
%	\left\{
%	\begin{array}{l}
%		\dot{x}_1 = x_2\\
%		\dot{x}_2 = \- 18.03 x_1 \- 4.51 x_2 \-8.76 x_3 \+ 10.99 u \+ 2.75 \dot{u} \-2.20 \rho(\dot{u}) \\
%		\dot{x}_3 = \rho(\dot{u})\+0.0125 x_3\\
%	\end{array}
%	\right.
%	\label{numbers}
%\end{equation}
%\normalsize
%
%%\section{\uppercase{Discussion}}
%
%\label{discussion}
\section{\uppercase{Conclusions and future work}}
\label{conclusions}
The formal analysis of the system has
provided a condition for the stability, 
specifically on the gains of the PD controller, 
i.e., \eqref{ineq1}-\eqref{ineq3}. This confirms the idea, suggested by empirical experiments with human subjects and robots,
that the nonlinearity is \textit{benign} in that it does not endanger the stability of the system. There is the hypothesis that such dead-band nonlinearity could be useful in cutting out vestibular noise, especially when the support surface is not moving (that is the most common scenario in nature). In order to study the effect of the threshold on noise future work may integrate methods for the analysis of stochastic systems \cite{ctde185,ctde181}. Another important aspect in posture control, that was not considered here, is the effect of delay. Delay imposes a limitation on feedback gain that can be considered the motivation for the feed-forward compensation of external disturbances (in this work, gravity). The effects of delay have been studied formally in the linear case \cite{antritter2014stability}, but not yet with the nonlinear system.  As the DEC has also been applied to multiple degrees of freedom scenarios \cite{lippi2019distributed,lippi2017human} the formal study may be extended to multiple inverted pendulum models. A way to tackle the complexity of the multiple DoF problem may require the use of numerical methods for the study of the stability \cite{ctde182,ctde183,ctde184}, this will require a particular effort considering the number of state variables required to represent the dynamics of the mechanical degrees of freedom, the dynamics of the sensory estimates (e.g. the leaky integrator in the presented work) and the ones used to represent the delays.
\vspace{-2mm}
\section*{\uppercase{Acknowledgements}}
\setlength{\intextsep}{-1pt}%
\setlength{\columnsep}{15pt}%
\begin{wrapfigure}{l}{0.08\columnwidth}
		{\includegraphics[width=0.12\columnwidth]{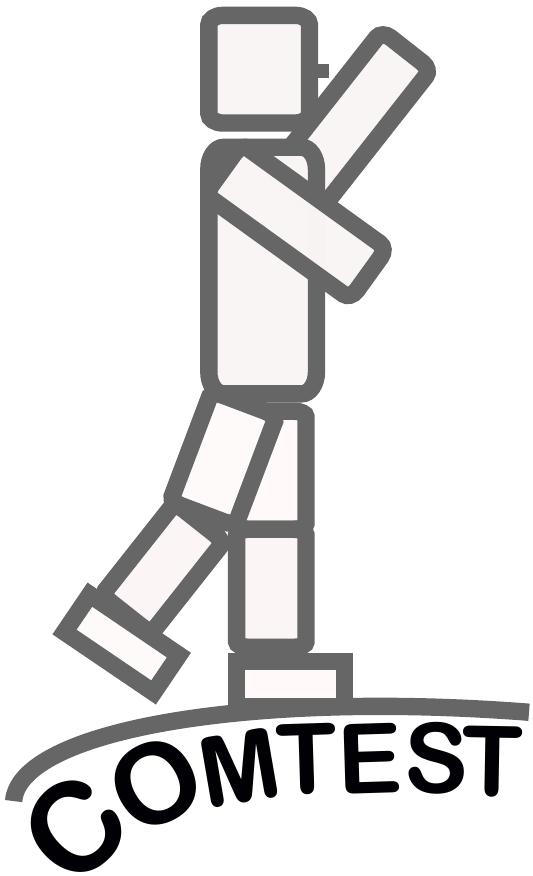}}
	\label{LOGO}
\end{wrapfigure}
\noindent This work is supported by the project COMTEST \cite{Lippi2019}, a sub-project of EUROBENCH (European Robotic Framework for Bipedal Locomotion Benchmarking, www.eurobench2020.eu) funded by H2020 Topic ICT 27-2017 under grant agreement number 779963.
\vspace{-2mm}

\bibliographystyle{apalike}
{\small
\bibliography{example}}

\end{document}

%% file: mathCommands.tex
\newcommand{\fa}[1]{\textcolor{black}{{#1}}}

\newcommand{\real}{\mathbb{R}}
\newcommand{\realp}{\mathbb{R}_{>0}}

%% file: bodyMechanics.tex
% !TeX root = Example.tex
\fa{
Human posture dynamics
in the sagittal plane is usually 
modelled as an inverted pendulum. 
Depending on the scenario,
the model can be a single inverted pendulum (SIP), see, e.g., \cite{Mergner2003,jafari2019stabilization},
or a multiple inverted pendulum, see, e.g., \cite{alexandrov2017human,Hettich2013,V.Lippi2013,lippi2017human,icinco12}. 
The number of degrees of freedom (DoF) 
representing the body dynamics 
is in general linked to the intensity 
of external stimuli, see \cite{atkeson2007multiple} for further details. 
In this work, we consider a SIP model that is 
used to represent the upright stance in 
presence of small disturbances. 
In literature, models with 2 DoF have
been used for modelling 
balance in the frontal plane \cite{goodworth2010influence,lippi2016human}.
\newline
In the remainder of this paper, we consider the following
scenario: the subject is balancing on a \textit{tilting platform} 
whose inclination is controlled by an external input. 
In order to keep the equilibrium despite the
tilting movement, the orientation in space of the inverted pendulum
is actively controlled by an ankle movement.
\newline
Formally, $\alpha_{BS}\in\real$ 
denotes the angle between the body (pendulum axis)
and the vertical (gravity axis).
$\alpha_{BF}\in\real$ denotes
the angle between the body and the axis
normal
to the tilting platform.
The third considered angle is $\alpha_{FS}\in\real$
that represents the angle of the tilting platform
with regards to the vertical axis.
From Figure \ref{fig:Fig1}, this three angles 
are linearly dependent, i.e.,
\begin{equation}
	\label{eq:angles_eq}
	\alpha_{BF}=\alpha_{BS}-\alpha_{FS}.
\end{equation}
The torque provided by the ankle 
is $T_a\in\real$, whereas the one
produced by the gravitational force is
$T_G\in\real$.
\newline
The pendulum dynamics is described
by 
\begin{equation}
	\ddot{\alpha}_{BS} = \frac{T_a+T_p+T_G}{J_B},
	\label{eqd1}
\end{equation}
where $J_B\in\realp$ 
is the moment of inertia of the body around the ankle joint
and $T_p\in\real$ is
the torque describing passive stiffness and damping, i.e.,
\begin{equation}
	T_{p}=K_p^{p}\alpha_{BF} + K_d^{p}\dot{\alpha}_{BF}
\end{equation}
The passive stiffness,
i..e, $K_p^p$, and damping,
i.e., $K_d^p$,
causes additional destabilisation. 
However, as in \cite{ott2016good},
they have a role in stabilising the system dynamics in presence of delays.
}

%% file: measuredSignals.tex
\subsection{Sensors' information}\label{sec:sensors}
\fa{
Signal $\check{\alpha}_{BS}\in\real$ provides
the estimate of $\alpha_{BS}$ (body sway) 
obtained by the vestibular system 
(or, for humanoids, by the IMU). 
On the other hand, signal $\check{\alpha}_{BF}\in\real$ is
the proprioceptive input measured at the ankle
(for humanoids, given by an encoder). 
Both signals are estimates
of the corresponding physical quantities.
%and are employed by the closed-loop
%controller.
Their derivatives with respect to time
are $\dot{\check{\alpha}}_{BS}\in\real$, 
sensed by the vestibular system,
and $\dot{\check{\alpha}}_{BF}\in\real$, 
sensed by the
proprioceptive system. 
\newline
In what follows, we put a check symbol 
above all measured
variables 
(i.e., $\check{x}$). On the other hand,
all estimated variables
have the "hat" symbol above them (i.e., $\hat{x}$).
%\newline
%The values $\check{\alpha}_{BS}$ and $\check{\alpha}_{BF}$
%are measures of the corresponding
%variable. 
%Similarly to (\ref{eq:angles_eq}),
%\begin{equation}
%	\label{eq:angles_eq_est}
%	\check{\alpha}_{BF}=\check{\alpha}_{BS}-\check{\alpha}_{FS},
%\end{equation}
%where $\check{\alpha}_{FS}\in\real$ is the estimate
%of the tilting platform angle, obtained by
%combining vestibular and proprioceptive
%senses.
}

%% file: humanInspiredSFC.tex
% !TeX root = Example.tex
\subsection{Gravity Compensation}
\fa{%
The gravity force is the largest effect acting on the body, see \cite{zebenay2015human}. 
Formally, the torque produced by this force
is 
\begin{equation*}
T_G=m_B \cdot g \cdot h_B \cdot \sin\left({\alpha}_{BS}\right),
\end{equation*}
where $g$ is the gravity acceleration,
$m_B\in\realp$ the body mass,
and $h_B\in\realp$ the height of the centre of mass. 
Under the assumption of a small angle,
\begin{equation}\label{eq:gravity}
	T_G\simeq m_B \cdot g \cdot h_B \cdot {\alpha}_{BS}.
\end{equation}
The ankle torque, actively produced by the subject
in order to keep the body standing despite the tilting platform,
i.e., $T_a$ in (\ref{eqd1}),
also compensates for this gravity disturbance.
In fact, let $T_a^G$ be the component
of $T_a$ compensating gravity, such that
%\hl{check if $T_a^a$ makes sense}
\begin{equation}
	T_a=-T_a^G + T_a^a,
\end{equation}
where $T_a^a$ is the ankle torque's component
not due to gravity compensation.
Gravity is slightly under-compensated in humans,
see, e.g., \cite{T.Mergner2009,G.Hettich2014},
thus, as in \cite{ott2016good}, we assume an
arbitrary gain, i.e., $K_G\in\realp$,
for gravity compensation. Thus,
\begin{equation}\label{eq:gravity_comp}
	T_a^G= K_G\cdot \check{\alpha}_{BS}.
\end{equation}
}
\subsection{Support Surface Tilt Compensation}
\label{sss}
\fa{%
In order to reproduce the
behaviour observed in humans, see, e.g., \cite{T.Mergner2009,Mergner2003,hettich2015human,G.Hettich2014},
the control input $T_a^a$ is not computed
by directly using
the measured quantity $\check{\alpha}_{BS}$,
but an estimate of ${\alpha}_{BS}$,
say $\hat{\alpha}_{BS}$.
To this end, first, 
the inspection of human
behaviour suggests to
use signal
$\dot{\hat{\alpha}}_{FS}\in\real$,
obtained
by using both vestibular and proprioceptive sensed values,
to estimate
the tilting platform's angle. 
Denote this estimate
by ${\hat{\alpha}}_{FS}\in\real$. This value is
then used for computing $\hat{\alpha}_{BS}$,
which closes the control loop. 
\newline
Formally, by (\ref{eq:angles_eq}),
one has
\begin{equation}
	\dot{\hat{\alpha}}_{FS}=\dot{\check{\alpha}}_{BS}-\dot{\check{\alpha}}_{BF}.
\end{equation}
%Note that, as explained in Section~\ref{sec:sensors},
%the checked variable denotes \hl{...}.
%As in \hl{[X]},
%the closed-loop control for keeping the stance
%is fed with 
The estimate of ${\hat{\alpha}}_{FS}$ 
simulates
the human behaviour. This is done by
feeding $\dot{\hat{\alpha}}_{FS}$ 
into function $\rho(\cdot)$,
which is then integrated through a 
leaky integrator, i.e.,
%
%The support surface rotation is estimated 
%by combining $\dot{\check{\alpha}}_{BS}$,
%sensed by the vestibular system (or IMU),
%with the ankle joint angle speed $\dot{\alpha_{BF}}$,
%sensed by the proprioceptive system (). 
%The tracking is based on speed in order to reproduce the nonlinear behavior observed in humans. A nonlinear \textit{threshold} function $\rho()$ in applied on the speed that is then integrated through a leaky integrator
\begin{equation}
	\hat{\alpha}_{FS} = \int^{t}_{0} \rho\left(\dot{\hat{\alpha}}_{FS} \right) -c_L{\hat{\alpha}}_{FS} d\tau
	\label{eq:leakyInt}
\end{equation}
with $c_L\in\realp$ and the threshold function
defined as
\begin{equation}\label{deadband}
	\rho(\alpha):= 
	\begin{cases}
		\alpha+\theta        &  \text{if }          \alpha\leq -\theta \\
		0                    &  \text{if }-\theta<   \alpha< \theta \\
		\alpha-\theta        &  \text{if }\theta \leq\alpha \\
	\end{cases},
\end{equation}
for $\theta\in\realp$.
\newline
With this piece of information at hand,
we compute $\hat{\alpha}_{BS}$,
the quantity used in $T_a^a$,
by employing (\ref{eq:angles_eq}),
i.e.,
\begin{equation}\label{eq:gettingalfahat}
	\hat{\alpha}_{BS}=\hat{\alpha}_{FS}+\check{\alpha}_{BF}.
\end{equation}
}
%To simulate what happens in humans,
%the estimate of
%body orientation affected by the 
%nonlinearity is then computed using 
%again the proprioceptive input, this time in terms of position to compute an up-channeled version
%\begin{equation}
%\hat{\alpha}_{BS}=\hat{\alpha}_{FS}+\alpha_{BF} = \alpha_{BS}+\hat{\alpha}_{FS}-\alpha_{FS}
%\end{equation}
%In case the noise affecting the sensory input is not considered then the nonlinearity can be applied directly on the physical variable $\dot{\alpha}_{FS}$.  
\subsection{Other Disturbances}
In order to completely describe the effect of the environment on the body, other disturbances should also be taken into account. Field forces can be produced, for example, by an horizontal translation of the support surface $x_{fs}$ leading to a torque $T_{trans} = \ddot{x}_{fs} \cdot h_B \cdot m_B$ and an external touch that can be estimated as $ T_{ext} = \ddot{\check{\alpha}}_{BS}J_B-T_a$. A robotic control applying also these disturbances is described in \cite{zebenay2015human}. Currently, a model of human support surface translation compensation is still object of research, and there are no evidences yet of a direct compensation of such disturbances. In this work only gravity will be considered.   

\subsection{Servo Control}
As in \cite{ott2016good},
the system is controlled through a $PD$ controller
with proportional coefficient, respectively derivative coefficient,
being $K_p^a\in\real$, respectively $K_d^a\in\real$, 
i.e.,
\begin{equation}
	T_a^a = K_p^a\epsilon + K_d^a  \frac{d\epsilon}{dt},
	\label{active}
\end{equation}
where the error variable $\epsilon$ is defined as
\begin{equation}
	\epsilon := \hat{\alpha}_{BS} - \alpha_\mathrm{ref},
\end{equation}
with the desired position being, in general,
$\alpha_\mathrm{ref} = 0$.
All delays involved into the processing of sensory inputs and the motor control are not considered in this analysis.
In neurology the concept was proposed in \cite{merton1953speculations} to explain the role of the muscle stretch reflex for the control of posture and movements: a PD-controller adjusts the force of the muscles so as to produce the desired pose or movement.
%A version with a $PID$ is also used in some works, with the effect of integrating the gravity torque and making the body stand upright from a leaning position. This behavior although observed in humans is not desirable for the control of humanoids where a desired position that is different from $0^\circ$ should be reached and kept. For this reason in some works the DEC has been implemented with an integrator that excluded the gravity and with also with a specific derivative coefficient for the disturbances \cite{ott2016good}.

\begin{remark}
	Gravity is compensated by directly using 
		the available measure $\check{\alpha}_{BS}$,
		whilst $T_a^a$ uses the estimated value $\hat{\alpha}_{BS}$.
		This is due to the fact that the nonlinearity has been experimentally observed 
		only on the latter \cite{T.Mergner2009,Mergner2003,G.Hettich2014}. Qualitatively, the effect on the threshold applied on $\hat{\alpha}_{BS}$ is that the slower the
		platform tilting is, 
		the less it is compensated (gain nonlinearity). 
		For very slow platform movements (i.e. $\dot{\alpha}_{FS}<\theta$), there is even no compensation. 
		A similar nonlinearity applied on the gravity compensation would produce a paradoxical behaviour.
\end{remark}

%% file: system.tex
% !TeX root = Example.tex

	\begin{ass}
		As done in \cite{lippi2017human,V.Lippi2013,mergner2010},
		we assume the direct measurements to be
		equal to the corresponding variables, i.e.,
		$\check{\alpha}_{BS}={\alpha}_{BS}$ and
		$\check{\alpha}_{BF}={\alpha}_{BF}$.
	\end{ass}
	Let the input to the system
	be 
	\begin{equation}
	u=\dot{\alpha}_{FS},
	\end{equation}
	i.e., the tilting speed of the platform.
	The state vector is four-dimensional and equal to
	\begin{equation}
	\mathbf{x}=
	\left[
	\begin{array}{c}
	x_1 \\
	x_2 \\
	x_3 \\	
	x_4 \\				
	\end{array}
	\right] = 
	\left[
	\begin{array}{c}
	\alpha_{BS} \\
	\dot{\alpha}_{BS} \\
	\hat{\alpha}_{BS} \\
	{\alpha}_{FS} \\
	\end{array}
	\right].
	\end{equation}
	Starting from (\ref{eqd1}),
	we derive a model for the system at hand,
	by incorporating (\ref{eq:gravity}), (\ref{eq:gravity_comp}), 
	(\ref{eq:leakyInt}), (\ref{eq:gettingalfahat}), and (\ref{active}).
	This yields the following system:
	\begin{align}\label{system}
	\begin{cases}
	\dot{x}_1
	&=x_2
	\\
	\dot{x}_2
	&=a_1x_1 + a_2x_2 + a_3x_3 + a_4x_4 + f(u)
	\\
	\dot{x}_3
	&=bx_1 + x_2 - bx_3 - bx_4 + g(u)
	\\
	\dot{x}_4
	&=u
	\end{cases}
	\end{align}
	where
	\begin{align}
	a_1 &= \cfrac{K_p^p+K_d^a c_L+mgh_B -K_G}{J_B}~,\label{a_1}\\
	a_2 &= \cfrac{K_d^a+K_d^p}{J_B}~,\\
	a_3 &= \cfrac{K_p^a-c_LK_d^a}{J_B}~,\\
	a_4 &= \cfrac{K_p^p-c_LK_d^a}{J_B}~,\\
	b   &= C_L~,\label{b}\\
	g(u)&= \rho(u)-u~,\\
	f(u)&= K_d^ag(u)-K_d^pu.
	\end{align}
	Note that the nonlinearity brought about
	by $\rho(\cdot)$ affects the system
	only through input $u$.
	Figure \ref{fig:fu} illustrates $f(u)$ and $g(u)$.
\begin{figure}[t]
	\input{fugu.tex}
	\caption{$f(u)$ and $g(u)$}
	\label{fig:fu}
\end{figure}

	System (\ref{system}) can be also written in matrix form\footnote{
		In this case we explicitly report the dependence on time.
	},
	i.e., 
	\begin{equation}
	\label{system_mat}
	\dot{\mathbf{x}}(t)
	=
	A\mathbf{x}(t) + B(u(t)),
	\end{equation}
	where
	\begin{equation}
	A:=
	\begin{bmatrix}
	0&1&0&0\\
	a_1&a_2&a_3&a_4\\
	b&1&-b&-b\\
	0&0&0&0
	\end{bmatrix}
	\end{equation}
	and
	\begin{equation}
	B(u):=
	\begin{bmatrix}
	0\\f(u(t))\\g(u(t))\\u(t)
	\end{bmatrix}.
	\end{equation}
	System (\ref{system_mat})
	is a linear system, with nonlinearity on the control input.
	The dynamics of the fourth state is a
	simple integrator of input $u$,
	thus $0$ is an eigenvalue of the
	system's dynamics.
	The remaining three eigenvalues can be
	determined by the choice of 
	$K_a^p$ and $K_a^d$, our
	design parameters. In the following the stability conditions are derived in an analytical way
	while in previous work the stability of the DEC was demonstrated empirically with simulations \cite{V.Lippi2013} and robot experiments \cite{G.Hettich2014,ott2016good,zebenay2015human}.
	\begin{lemma}\label{lemma1}
		If $K_p^a$ and $K_d^a$ are chosen
		such that
		\begin{align}
		K_d^a&<c_LJ_B-K_d^p,\label{ineq1}\\
		K_p^a+K_d^a c_L &< K_G-mgh_B - K_p^p - c_Lk_d^p,\label{ineq2}\\
		K_p^a&<K_G-mgh_B - K_p^p,\label{ineq3}
		\end{align}
		then system~(\ref{system_mat})
		has three eigenvalue with negative real part.
		\begin{proof}
			By definition of eigenvalues,
			the spectrum of $A$
			is
			$$
			\mathrm{eig}(A) = 
			\{0\}
			~\cup~
			\mathrm{eig}(\tilde{A}),
			$$
			where
			$$
			\tilde{A}:=
			\begin{bmatrix}
			0 &1 &0\\a_1 &a_2 &a_3\\
			b &1 &-b
			\end{bmatrix}.
			$$
			The characteristic polynomial of $\tilde{A}$,
			whose solutions are $\tilde{A}$'s eigenvalues,
			is 
			$$
			p_A(\lambda)
			=
			\lambda^3
			+
			(b-a_2)\lambda^2
			+
			(-a_3-a_2b-a_1)\lambda-b(a_1+a_3).
			$$
			By the Descartes \textit{Rule of Signs},
			we can impose that all eigenvalues of
			$\tilde{A}$ have negative real part, by
			holding
			$$
			\begin{cases}
			b-a_2>0\\
			-a_3-a_2b-a_1>0\\
			-b(a_1+a_3)>0
			\end{cases}.
			$$
			This latter becomes
			a set of inequalities in
			$K_d^a$ and $K_p^a$,
			by incorporating (\ref{a_1})-(\ref{b}).
			This yields (\ref{ineq1})-(\ref{ineq3}),
			thus concluding the proof.
		\end{proof}
	\end{lemma}
	\begin{lemma}\label{lemma2}
		The solution to system~(\ref{system_mat})
		is
		\begin{equation}\label{eq:sol}
		{\mathbf{x}}(t)
		=
		e^{At}{\mathbf{x}}(0) + 
		\int_{0}^{t}
		e^{A(t-\tau)}
		B(u(\tau))d\tau.
		\end{equation}
		\begin{proof}
			Let $u_1(t):=f(u(t))$,
			$u_2(t):=g(u(t))$,
			and $u_3(t):=u(t)$.
			We have
			\begin{equation}\label{eq:equivalentLinSys}
			\dot{\mathbf{x}}(t)
			=
			A\mathbf{x}(t) + \tilde{B}\mathbf{\tilde{u}}(t),
			\end{equation}
			with
			\begin{equation}\label{eq:Bchange} 
			\tilde{B}\mathbf{\tilde{u}}(t)
			=
			B(u(t)) 
			\end{equation}
			where 
			$
			\mathbf{\tilde{u}}(t) :=
			[u_1(t),u_2(t),u_3(t)]'
			$ and
			$$
			\tilde{B} :=
			\begin{bmatrix}
			0 &0 &0\\
			1 &0 &0\\
			0 &1 &0\\
			0 &0 &1
			\end{bmatrix}.
			$$
			By \cite[(4.7)]{skogestad2007multivariable},
			$$
			{\mathbf{x}}(t)
			=
			e^{At}{\mathbf{x}}(0) + 
			\int_{0}^{t}
			e^{A(t-\tau)}
			\tilde{B}\mathbf{\tilde{u}}(\tau)d\tau,
			$$
			which, by incorporating~(\ref{eq:Bchange}),
			yields~(\ref{eq:sol}), thus concluding the proof.
		\end{proof}
	\end{lemma}
	In (\ref{eq:sol}), the first addendum
	is the free response, and the integral is 
	referred to as forced response. 
	By~(\ref{eq:sol}),
	system's stability is determined only by
	matrix $A$, thus the nonlinearity
	acting on the input, i.e., $B(u(t))$,
	does not play any role for stability.
	The following two definitions of stability
	are extracted from \cite[Chapter 3]{mellodge2015practical}
	and \cite{bernstein1995lyapunov}.
	\begin{definition}[Asymptotic Stability]\label{def:as} 
		System~(\ref{eq:equivalentLinSys}) is 
		asymptotically stable if and only if
		all the eigenvalues of 
		$A$ are in the left half of the complex plane.
	\end{definition}
	\begin{definition}[Lyapunov Stability]\label{def:ls} 
		System~(\ref{eq:equivalentLinSys}) is 
		Lyapunov stable if and only if 
		no eigenvalues of A are 
		in the right half of the complex plane
		and all eigenvalues on the imaginary axis
		are semisimple (i.e., they
		have algebraic multiplicity equal
		to the geometric multiplicity).
	\end{definition}
	\begin{thm}
		If $K_p^a$ and $K_d^a$ are chosen
		as in Lemma~\ref{lemma1},
		system~(\ref{system_mat})
		is Lyapunov stable, but not asymptotically stable.
		\begin{proof}
			Consider system~(\ref{eq:equivalentLinSys})
			which is an equivalent of~(\ref{system_mat}).
			Clearly,
			system (\ref{eq:equivalentLinSys}) is Lyapunov stable (asymptotically stable)
			if 
			and only if
			also (\ref{system_mat}) is Lyapunov stable (asymptotically stable).
			\newline
			By Lemma~\ref{lemma1}, matrix
			$A$ has three eigenvalues
			with negative real part and one eigenvalue
			(the integrator in $x_4$)
			which is on the imaginary axis and semisimple. 
			By Definition~\ref{def:ls},
			system~(\ref{eq:equivalentLinSys}) is
			Lyapunov stable.	
			By Definition~\ref{def:ls},
			system~(\ref{eq:equivalentLinSys})
			is not asymptotically stable, thus
			the proof is concluded.
		\end{proof}
	\end{thm}
The non-linear system can be stabilised by choosing the appropriate pair of proportional and derivative coefficients for the PD controller.

%% file: fugu.tex
% !TeX root = Example.tex
\usetikzlibrary{decorations.text}
\begin{tikzpicture}
%% AXIS
\begin{axis}[%
	axis lines*=middle,
	width=\columnwidth,
	height=.5\columnwidth,
	scale only axis,
	xmin=-5,
	xmax=5,
	xtick={-2.5,  2.5},
	xticklabels={$-\theta$, $\theta$},
	ymin=-4,
	ymax=4,
	ytick={-2.5,  0,  2.5},
	yticklabels={$-\theta$, $ $, $\theta$}],
% g(u)
\addplot [dashed, red, thick, postaction={decorate,decoration={text along path, raise=2pt,
		text={g(u)}}}, domain=-5:-2.5] {2.5};
\addplot [dashed, red, thick, domain=2.5:5] {-2.5};
\addplot [dashed, red, thick, domain=-2.5:2.5] {-x};
% f(u)
\addplot [dashed, blue, thick, domain=-5:-2.5] {-0.2*x+3};
\addplot [dashed, blue, thick, postaction={decorate,decoration={text along path, raise=1pt,
		text={f(u)}}}, domain=2.5:5] {-0.2*x-3};
\addplot [dashed, blue, thick, domain=-2.5:2.5] {-3.5/2.5*x};
\end{axis}
\end{tikzpicture}